# Chemosensitivity testing of revived fresh-frozen biopsies using digital speckle holography

Zhen Hua[1], John Turek[2], Mike Childress[3], and David Nolte[1]

[1]*Department of Physics and Astronomy, Purdue University, 525 Northwestern Ave, West Lafayette, Indiana 47907, USA*
[2]*Department of Basic Medical Sciences, Purdue University, 625 Harrison Street, West Lafayette, Indiana 47907, USA*
[3]*Department of Veterinary Clinical Sciences, Purdue University, 625 Harrison Street, West Lafayette, Indiana 47907, USA*

Enrolling patients in clinical trials to obtain fresh tumor biopsies to profile anticancer agents can be slow and expensive. However, if flash-frozen biopsies can be thawed to produce viable living tissue with relevant biodynamic profiles, then a large reservoir of tissue-banked samples could become available for phenotypic library building. Here, we report biodynamic profiles acquired from revived flash-frozen canine B-cell lymphoma biopsies using digital speckle holography. We compared the thawed-tissue drug-response spectrograms to spectrograms from fresh tissues in a study of canine B-cell lymphoma. By compensating for tissue trauma in the thawed sample, patient clustering of both the fresh and thawed samples were found to be in general agreement with clinical outcomes. This study indicates that properly frozen tumor specimens are a viable proxy for fresh specimens in the context of chemosensitivity testing, and that thawed samples from tissue banks contain sufficient viable cells to evaluate phenotypic drug response.

## 1 Introduction

### 1.1 Digital holographic optical coherence imaging

Biodynamic profiling is an optical imaging technology related to *en face* OCT [1] using partially coherent speckle generated by broad-area illumination with coherence detection through digital holography [2]. Biodynamic profiling penetrates up to 1 mm into living tissue and returns high-content information in the form of dynamic light scattering across a broad spectral range. The fluctuation frequencies relate to Doppler frequency shifts caused by light scattering from subcellular constituents that are in motion [3]. The speeds of intracellular dynamics range across nearly four orders of magnitude from nanometers per second (cell membrane motion) to tens of microns per second (organelles and vesicles movement). For a near-infrared backscattering geometry these speeds correspond to Doppler frequencies from 0.01 Hz to 10 Hz. Dynamic light scattering in living tissues has been used to identify intracellular transport signatures of diffusive relative to directed motion [4], for the detection of apoptosis [5], and extracellular restructuring [6].

### 1.2 Intracellular dynamics of living tissue

Intracellular dynamics in living tissue are dominated by active transport driven by bioenergetic processes far from thermal equilibrium [7]. Cells are highly dynamic systems that continuously undergo internal reconfiguration through random and/or coordinated molecular and mechanical responses. Intracellular dynamics are fundamental processes that support a broad range of functions such as cell migration and division [8]. These intracellular processes are derived from, and often influence, physiological conditions of the cells. Quantitative measurement of intracellular processes would thus aid in building a better understanding of the underlying mechanisms of cellular states and functions.

Dynamic light scattering combined with coherence-gated optical sectioning has led to the development of biodynamic imaging (BDI) [9] and related techniques such as tissue dynamics spectroscopy (TDS) [10]. Biodynamic profiling techniques are sensitive probes of the response of living tissue to applied drugs and therapeutics [9, 11], which has been extended to profiling how living biopsy samples respond to standard-of-care anticancer treatments. Biodynamic studies of chemosensitivity in patients obtain living biopsy samples through a conventional diagnostic process. However, a canine B-cell lymphoma and a human ovarian cancer trial required several years to enroll approximately 20 in each study [12, 13]. This slow rate of enrollment limits the numbers of samples that can be obtained. To identify biodynamic signatures of drug sensitivity or resistance in the face of sample-to-sample and

---
*Corresponding author*
*e mail:* nolte@purdue.edu (David Nolte)

patient-to-patient heterogeneity requires phenotypic profiles of at least 50 independent samples depending on the variability of the biodynamic spectral fingerprints.

If flash-frozen biopsies could be revived and measured, then a large reservoir of tissue-banked samples could become available for phenotypic library building. In this paper, we demonstrate that fresh-frozen biopsy samples can be thawed, and their health stabilized sufficiently to measure biodynamic spectral signatures of their responses to applied therapeutics. Two tissue types are studied here: canine B-cell lymphoma, for which we have both fresh and frozen tissues as well as the patient clinical outcomes. Biodynamic intracellular processes occur in the thawed tissues that do not match fresh tissue, but these effects can be partially compensated to allow a comparison between fresh and thawed tissues in the case of the canine lymphoma.

## 2 Biodynamic profiling system and experimental methods

### 2.1 Biodynamic profiling system

The experimental setup of the biodynamic imaging system is shown in Fig 1 based on a Mach-Zehnder interferometer and off-axis digital holography. The bandwidth of the light source (Superlum, Cork, Ireland) is 50 nm, the wavelength is 840 nm, and the coherence length is ~15 µm. The scattering from the specimen serves as the signal while the reflection from the first beam splitter serves as the reference arm. The crossing angle between the reference beam and the signal beam is two degrees and can be changed by tuning the orientation of the final beam splitter or the optical path delay (mirror system). A neutral density (ND) filter is placed on the reference arm to reduce the intensity of the reference. The CCD camera is placed on the focal plane of the third lens.

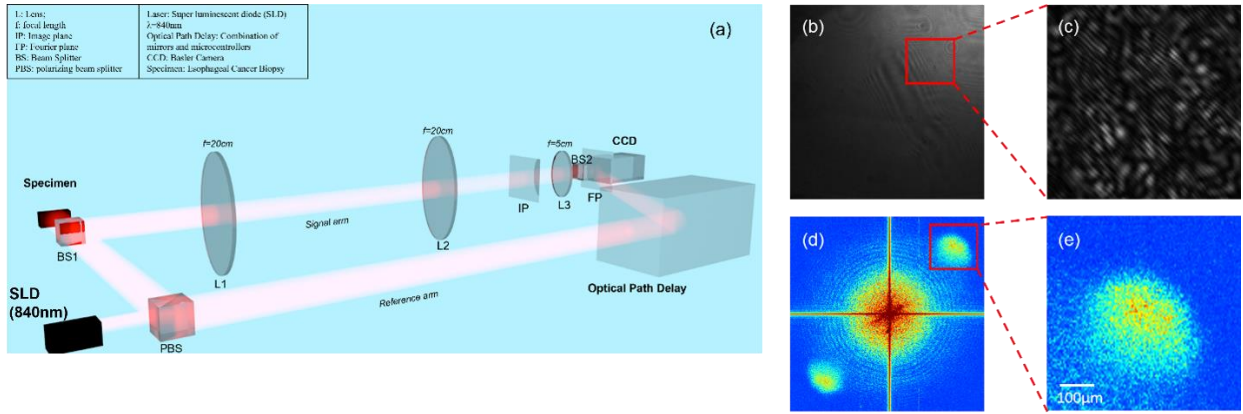

Fig 1. Experimental Setup and Dataflow. (a) Optical coherence imaging (OCI) system configuration with low-coherence light. (L1-3: lenses. FP: Fourier plane. IP: image plane. CCD: charge-coupled device digital camera). (b) A single hologram captured on the camera plane, which is also a FP. (c) the blow-up of hologram. (d) the Optical coherence image (OCI) reconstruction and its phase conjugate produced by a two-dimensional fast Fourier transform (FFT). (e) the blow-up of OCI reconstruction.

### 2.2 power spectrogram format

A sequence of OCI frames is captured, representing one observation set of the living target. By capturing several sequences, changes in the time-dependent behavior of the target are detected over many hours. For a sequence of OCI frames, the temporal normalized standard deviation (NSD) of the intensity $I$ is defined at each pixel (x,y) as

$$NSD(x,y) = \frac{\sigma_I(x,y)}{\langle I(x,y) \rangle} = \frac{\sqrt{[\langle I^2 \rangle - \langle I \rangle^2]}}{\langle I \rangle} \quad (1)$$

The NSD map is called the motility contrast image (MCI). Different biological processes happen at characteristic speeds. All these processes result in local fluctuations in the index of refraction of cells and tissues and cause dynamic changes in the scattered speckle. The autocorrelation of the intensity I of a pixel is given by

$$A(\tau) = \langle I(0)\, I(\tau)\rangle = \langle I\rangle^2 + [\langle I^2\rangle - \langle I\rangle^2]\, exp\left\{-\frac{\tau}{\tau_C}\right\} \quad (2)$$

where $\tau_C$ is the correlation time of the process. For diffusion and backscatter,

$$1/\tau_C = q^2 D = 4Dk_i \quad (3)$$

with $k_i$ being the magnitude of the wavevector of the incident light in the medium and D being the diffusion coefficient. The autocorrelation can be written as

$$A(\tau) = 1 + (NSD)^2\, exp\left\{-\frac{\tau}{\tau_C}\right\} \quad (4)$$

The Fourier transform is a Lorentzian

$$S(\omega) = \frac{(NSD)^2\, \tau_C}{(\omega\, \tau_C)^2 + 1} \quad (5)$$

In log-frequency space, $S(\omega)$ has a distinct shape and exhibits a knee frequency at

$$\omega_C = \frac{1}{\tau_C} = q^2 D \quad (6)$$

The observed process in living tissues is not strictly diffusive, presenting evidence of Levy statistics and heavy tails [16] and hence the power spectrum can be approximated as

$$S(\omega) = \frac{(NSD)^2/\omega_C}{(\omega\,/\omega_C)^s + 1} \quad (7)$$

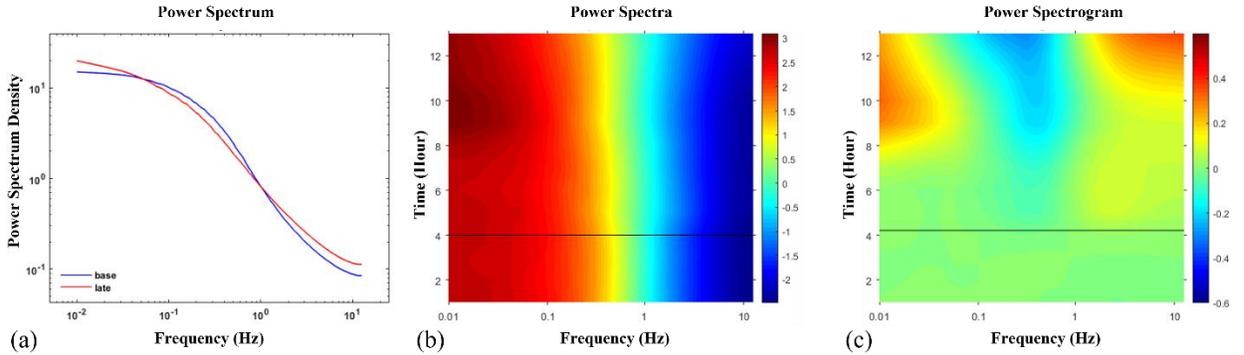

Fig 2. Spectra and spectrograms. (a), The power spectrum is the sum of all motion processes inside the tissues. (b), Fluctuation power spectra are acquired by performing temporal FFTs over reconstructed time series of optical coherence images of dynamic speckle monitored repeatedly over many hours. (c), The spectrogram is defined as the time dependent difference of the spectra between the baseline and the post-dose measurement. X axis is frequency, Y axis is time. The baseline measurement (prior to drug application) occurs in the first 4 hours. Post-drug-application measurement spans 10 hours

*2.3 Tissue dynamics spectroscopy*

The central data format of biodynamic profiling of intracellular dynamics inside living tissue is the drug-response spectrogram defined as

$$D(\omega, t; r) = logS(\omega, t; r) - logS_0(\omega, t_0; r) \quad (8)$$

where S (ω, t; r) is the spectral power density at time t for the voxel located at r = (x, y, z). The spectrogram is referenced to the baseline at time $t_0$ prior to the application of the drug. Spectrograms are typically taken at a fixed depth z (usually near the midsection of the biopsy) and averaged over (x, y) to yield an average spectrogram for the sample.

## 3 Materials and methods

Original data on Fresh samples were collected from canine B-cell lymphoma patients. The common treatment applies a CHOP regimen therapy, which is a combination of four different cancer drugs (doxorubicin, cyclophosphamide, prednisolone, vincristine). Progression-free survival (PFS) defines the chemotherapy response sensitivity (Table 1). Progression-free survival is defined as the length of time during and after the treatment that a patient lives with the disease without progression. A PFS longer than 180 days is considered as a chemotherapy sensitive response for the canine patients.

The frozen tissues were snap-frozen in liquid nitrogen within 10-15 minutes of collection from the animal. All the tissue samples were kept frozen in liquid nitrogen in a large tank in the biorepository until the time of thaw/use. Upon retrieval, the samples were rapidly thawed by agitation in a 37ºC water bath and suspended in 37ºC RPMI medium containing 10% fetal bovine serum and 100U penicillin/mL-100µg/mL streptomycin. Small 1 $mm^3$ pieces were then assayed as previously described. For each day's experiment sixteen canine B-cell lymphoma samples are placed in a 96-well plate. There are four negative-control wells treated with 0.1% DMSO carrier and twelve wells treated with the drugs with duplicates (CHOP, prednisolone, vincristine) and triplicates (doxorubicin, cyclophosphamide).

Table 1. Progression-free survival (PFS) is defined as the length of time during and after the treatment of a disease, such as cancer, that a patient lives with the disease without progression. PFS longer than 180 days is defined as the chemotherapy response sensitive group. From this table, canine patients Kodie, Graciebell, Sinker and Chester are in the drug sensitive group, and canine patients Badger, Scout, Boost are in the drug resistant group. DLBCL: Diffuse large B-cell lymphoma; CR: complete remission; PR: partial remission; OS: overall survival. [1] only completed half of CHOP protocol.

Patient Profile

| Dog # | Dog name | Date of biopsy | Date of thawed assay | Histologic diagnosis | Best overall response | PFS | OS | Chemotherapy response |
|---|---|---|---|---|---|---|---|---|
| 1 | Kodie | 4/15/2015 | 10/24/2017 | DLBCL | CR | 223 | 244 | Sensitive |
| 2 | Graciebell | 5/14/2015 | 9/13/2017 | DLBCL | CR | 384 | 384 | Sensitive |
| 3 | Badger | 6/4/2015 | 12/4/2017 | DLBCL | PR | 64 | 138 | Resistant |
| 4 | Mr. Sinker[1] | 6/11/2015 | 9/29/2017 | DLBCL | CR | 460 | 460 | Sensitive |
| 5 | Scout | 3/23/2016 | 4/4/2018 | DLBCL | PR | 43 | 58 | Resistant |
| 6 | Chester | 5/6/2016 | 4/9/2018 | DLBCL | CP | 244 | 337 | Sensitive |
| 7 | Boost | 6/16/2016 | 4/6/2018 | DLBCL | CP | 98 | 156 | Resistant |

## 4 Biodynamic profiling of canine B-Cell biopsies

Biodynamic time–frequency spectrograms $D(\omega, t)$ were acquired for the seven enrolled fresh dogs under four treatments (doxorubicin, prednisolone, vincristine and cyclophosphamide) as well as wells dosed only with the 0.1% DMSO carrier as negative controls. Each treatment was measured using four or six replicates in randomized well locations. The averaged spectrograms of the negative control are shown in Fig 3 for fresh, thawed, and their difference, respectively. The spectrogram of the fresh samples shows a progressive inhibition of activity for low and mid frequencies. The decrease in activity with time is often associated with naturally decreasing sample health. The spectrogram for the thawed samples shows a strong inhibition across the full bandwidth. This stronger suppression is caused by the tissue damage associated with the freeze-thaw process. The difference of the thawed DMSO spectrogram relative to the fresh shows strongest relative suppression in high and low frequencies for the thawed tissues.

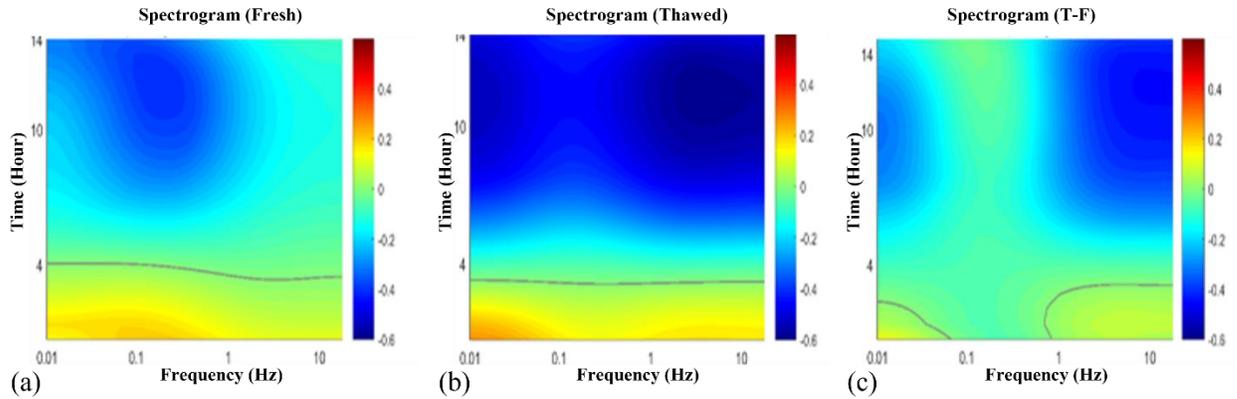

Fig 3. Negative control (DMSO-based response) spectrograms. X-axis is frequency; Y-axis is measurement time. The 0.1% DMSO in growth medium is applied after 4 hours. **(a)** Spectrograms for fresh biopsy, **(b)** Spectrograms for thawed sample, **(c)** The difference of spectrogram between thawed samples and fresh biopsies.

The averaged drug-response spectrograms for the Fresh and Thawed samples are shown in Fig 4. These spectrograms have had the average negative control response subtracted. They are arranged in four groups according to clinical outcomes: 1) Fresh drug-Resistant group, 2) Thawed drug-Resistant group, 3) Fresh drug-Sensitive group, 4) Thawed drug-Sensitive group for the four drug treatments. When comparing the trend for overall spectral response between the Fresh and Thawed cohorts with respect to the chemotherapy response phenotype, there are notable differences between the Fresh and Thawed spectrograms.

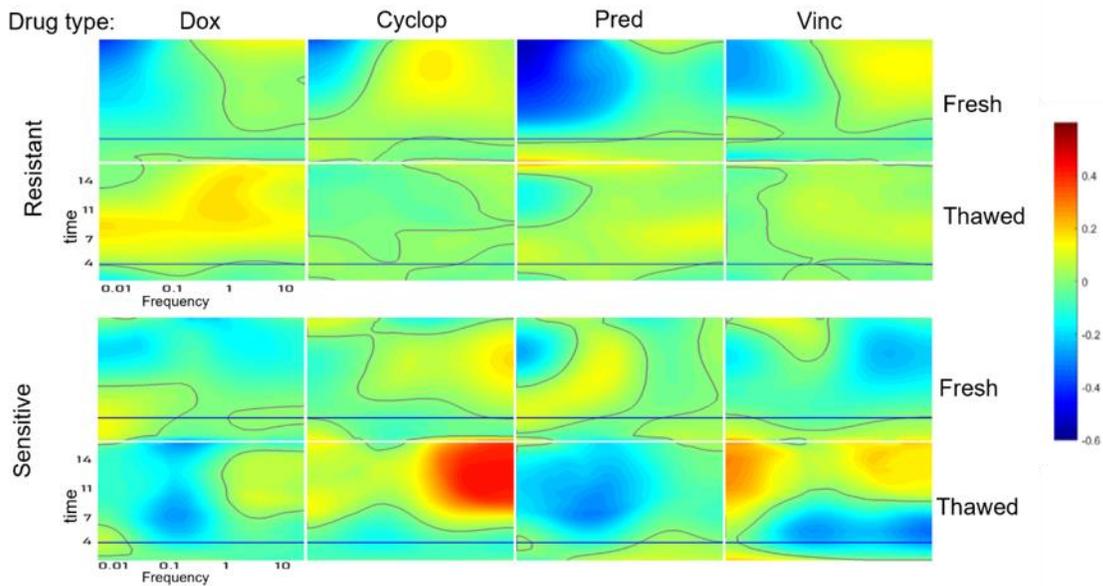

Fig 4. Relative spectrogram comparison of Fresh/Thawed and drug Sensitive/Resistant for the BDI results. Each column stands for a relative spectrogram (negative control removed) comparison based on one specific drug response under four different conditions: Fresh drug-Resistant biopsy, Thawed drug-Resistant biopsy, Fresh drug-Sensitive biopsy, Thawed drug-Sensitive biopsy. X-axis is Frequency, Y-axis is measurement time. Drug has been applied after 4 hours. All the drug response spectrograms have already subtracted the corresponding negative control response spectrogram.

The differences between the averaged Thawed relative to the Fresh spectrograms are shown in Fig 5. These figures show a rough correspondence of spectrograms within each cohort. For instance, all the drug-Resistant spectrograms show a strong enhancement in low frequency, while all the drug-Sensitive spectrograms show a strong enhancement in high frequency. These data show a consistent trend for overall enhanced spectral responses in short-PFS phenotypes relative to long-PFS phenotypes when under treatment.

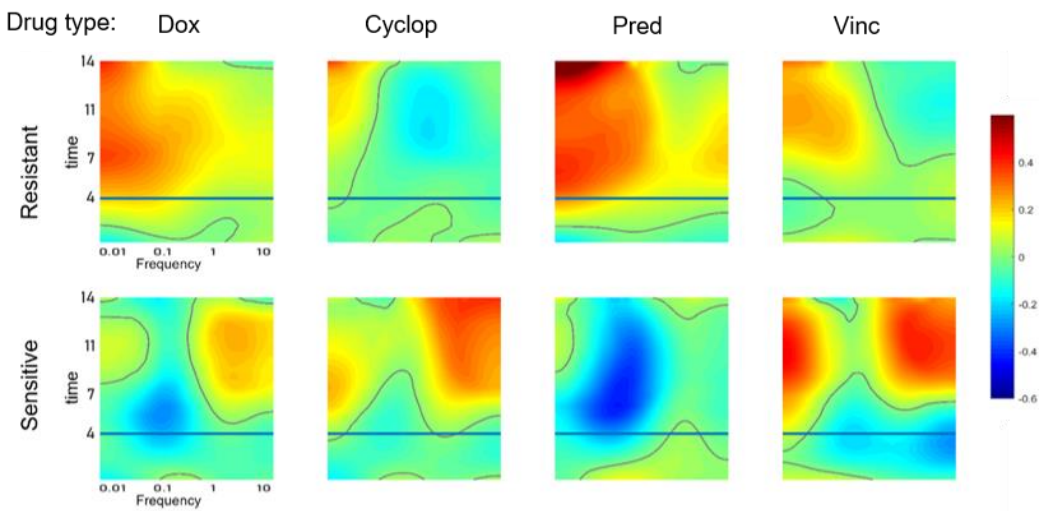

Fig 5. Relative spectrograms of Thawed groups relative to fresh drug responses grouped by cohort. The relative spectrograms are generated using the spectrogram of Thawed samples subtracting the spectrogram of corresponding Fresh biopsies grouped within the same chemotherapy and cohort. X-axis is Frequency and Y-axis is measurement time. Drug is applied after 4 hours.

**5 Machine learning and data clustering**

The central goal of biodynamic phenotyping is to establish libraries of spectral fingerprints for living tissue response to chemotherapies that show different phenotypes between patients who are sensitive or resistant to therapy. These spectral libraries can then be used in machine-learning classifiers to predict whether prospectively-enrolled patients will be sensitive or resistant to a selected chemotherapy. Because fresh tissue samples are relatively expensive to acquire, the ability to build libraries from frozen tissues in tissue banks would be a significant resource, if the tissue damage caused by the freeze/thaw can be characterized and removed.

A key goal of this preclinical study is to construct a chemoresistance classifier that takes a set of treatment spectrograms for a single patient and predicts whether the patient will have a sensitive response to a selected treatment regimen. To accomplish this, the drug-response spectrograms for each patient are deconstructed into a set of mathematical features, each capturing either local or global patterns. The construction of the classification algorithm is based on linear separability in a feature space. The time-frequency drug-response spectrograms for each patient are deconstructed into a set of features, each capturing either local or global spectrogram patterns. Examples include overall enhancement/suppression over all frequencies (ALLF, ALLFT); localized low, mid and high (HI) frequencies; red shifts or blue shifts (SDIP); and different time dependences in response to the applied treatment (SDIP vs. SDIP2 and ALLF vs. ALLFT). Over 40 biodynamic feature vectors are defined, described in previous publications on biodynamic profiling [9,14,15].

Fig 6 shows a combination result of all Fresh and Thawed samples. Many of the patients from the drug-Sensitive group are clustered with the drug-Resistant group. The classification accuracy is low, as expected because of the freezing/thawing trauma.

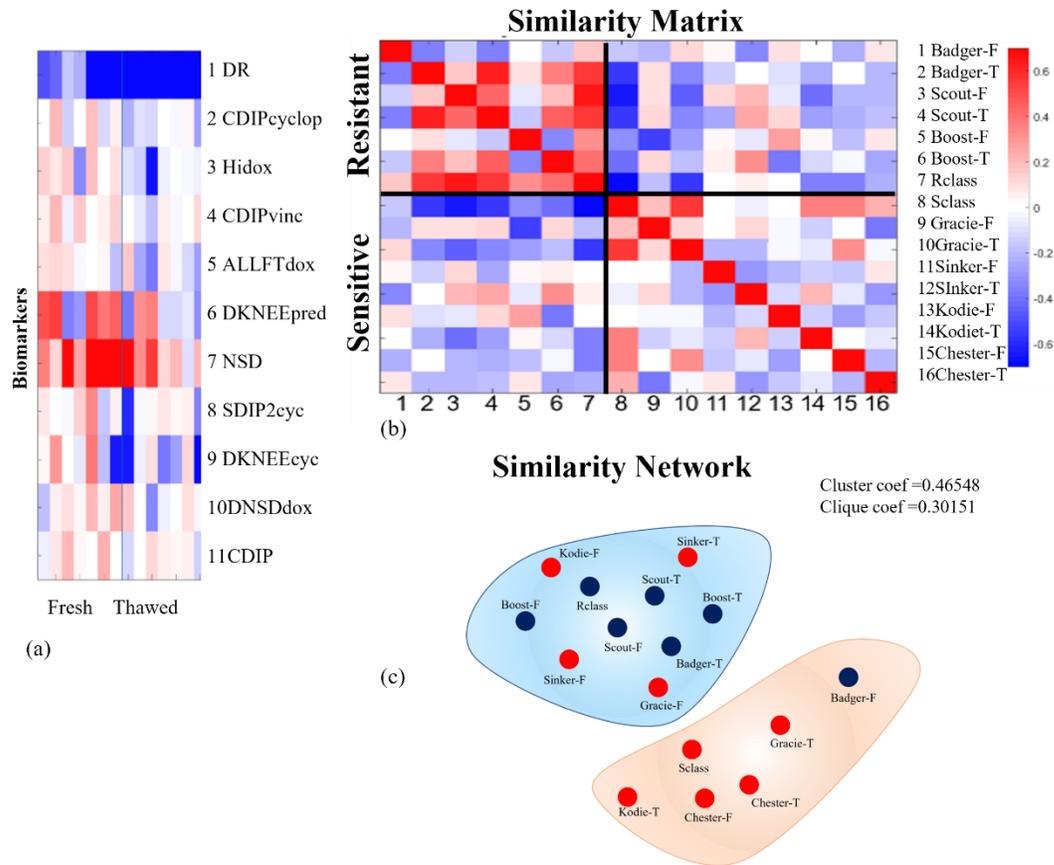

Fig 6. Clustering analysis. (a) Selected biomarkers, which are the most important features for distinguishing the two different cohorts. (b) Clustered Similarity matrix for all dogs (1-7 represents Resistant groups, 8-16 represents the Sensitive groups). Each grid stands for the correlation coefficient of 32 biomarkers between the two corresponding dog samples. The matrix shows a weak block diagonal pattern with two main groups. (c) Similarity network for all dogs, blue dots stand for drug-response resistant cohort, while red dots present drug-response sensitive cohort. Clique coefficient is defined as the accuracy of clustering into the correct group.

The thawed trauma can be partially eliminated by subtracting the averaged excess negative control response. Fig 7 shows the result of all Fresh and Thawed samples after the compensation. Overall, 12 out of 14 samples are classified correctly. These clustering results show a clearer distinction between the different chemotherapy response groups. It also shows that the compensation can partially remove the excess effect of the thawing damage.

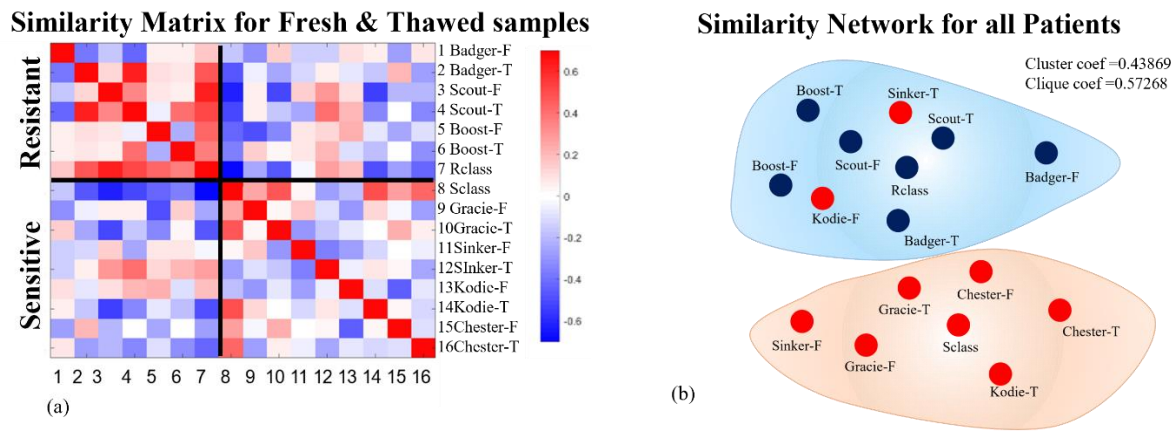

Fig 7. Similarity matrix after freezing trauma compensation. **(a)** Similarity matrix for all dogs after compensation (1-7 stands for Resistant groups, 8-16 stands for Sensitive groups). Each grid stands for the correlation coefficient of 32 biomarkers between the two corresponding dog samples. The matrix shows a block diagonal pattern with two main groups. **(b)** Similarity network for all dogs after compensation, blue dots are Resistant groups, red dots are Sensitive groups. Clique coefficient is defined as the accuracy of clustering into the correct group.

**6 Discussion & Conclusion**

The challenge for many clinical drug studies is the time necessary (sometimes months to years) to follow patient response to determine effectiveness. The use of frozen banked tissue with known clinical outcome represents a possible source of material for testing, provided enough cells survive the freezing process, and they maintain their phenotype. The goal of this project was to determine if thawed samples from neoadjuvant patients could be used to accurately assess drug response phenotype. Many cells in a thawed sample will be damaged due to ice crystal formation and will not survive the thaw process. However, some percentages do survive and cells can be grown out from thawed tissues. In addition to the problem of cryo-damage to the cells, there is also the question of whether the surviving cells would maintain the drug response phenotype as noted in the pathology report.

This study demonstrates that sufficient viable cells exist in a thawed sample to assess drug-response phenotype using biodynamic profiling. The relatively short revival time for biodynamic profiling analysis may be essential for this success. A significant advantage of biodynamic profiling is that the tissue can be monitored in the multiwell plate relatively soon after thawing. A more prolonged processing and/or culture time may not provide similarly consistent data since degradative processes (apoptosis, necrosis) can be triggered by the freeze-thaw stress.

**Acknowledgements**

This research was supported by NSF CBET-2200186.

**References**
1. Thouvenin O, Grieve K, Xiao P, Apelian C, and Boccara A C, *En face* coherence microscopy Invited, *Biomedical Optics Express*, 2017 doi.org/10.1364/BOE.8.000622
2. Jeong K, Turek J, and Nolte D, Speckle fluctuation spectroscopy of intracellular motion in living tissue using coherence-domain digital holography, *Journal of Biomedical Optics*, 2010 doi.org/10.1117/1.3456369
3. Li Z, Sun H, Turek J, Jalal S, Childress M, and Nolte D, Doppler fluctuation spectroscopy of intracellular dynamics in living tissue, *Journal of the Optical Society of America a-Optics Image Science and Vision*, 2019 doi.org/10.1364/JOSAA.36.000665
4. Lee J, Wu W C, Jiang J Y, Zhu B., and Boas D A, Dynamic light scattering optical coherence tomography, *Optics Express*, 2012 doi.org/10.1364/OE.20.022262
5. Farhat G, Giles A, Kolios M C, and Czarnota G J, Optical coherence tomography spectral analysis for detecting apoptosis in vitro and in vivo, *Journal of biomedical optics*, 2015 doi.org/10.1117/1.JBO.20.12.126001
6. Blackmon R L, Sandhu R, Chapman B S, Casbas-Hernandez P, Tracy J B, Troester M A, and Oldenburg A L, Imaging Extracellular Matrix Remodeling In Vitro by Diffusion-Sensitive Optical Coherence Tomography, *Biophysical Journal*, 2016 doi.org/10.1016/j.bpj.2016.03.014
7. Needleman D and Dogic Z, Active matter at the interface between materials science and cell biology, *Nature Reviews Materials*, 2017 doi.org/10.1038/natrevmats.2017.48

8. Gerencser A A and Nicholls D G, Measurement of instantaneous velocity vectors of organelle transport: Mitochondrial transport and bioenergetics in hippocampal neurons, *Biophysical Journal*, 2008 doi.org/10.1529/biophysj.108.135657

9. An R, Merrill D, Avramova L, Sturgis J, Tsiper M, Robinson J P, Turek J, and Nolte D, Phenotypic Profiling of Raf Inhibitors and Mitochondrial Toxicity in 3D Tissue Using Biodynamic Imaging, *Journal Of Biomolecular Screening*, 2014 doi.org/10.1177/1087057113516674

10. Nolte D, An R, Turek J, and Jeong K, Tissue dynamics spectroscopy for phenotypic profiling of drug effects in three-dimensional culture, *Biomedical Optics Express*, 2012 doi.org/10.1364/BOE.3.002825

11. Sun H, Merrill D, An R, Turek J, Matei D, and Nolte D, Biodynamic imaging for phenotypic profiling of three-dimensional tissue culture, *Journal of Biomedical Optics*, 2017 doi.org/10.1117/1.JBO.22.1.016007

12. Li Z, An R, Swetzig W M, Kanis M, Nwani N, Turek J, Matei D, and Nolte D, Intracellular optical doppler phenotypes of chemosensitivity in human epithelial ovarian cancer, *Scientific Reports*, 2020 doi.org/10.1038/s41598-020-74336-x

13. Choi H, Li Z, Sun H, Merrill D, Turek J, Childress M, and Nolte D, Biodynamic digital holography of chemoresistance in a pre-clinical trial of canine B-cell lymphoma, *Biomedical Optics Express*, 2018 doi.org/10.1364/BOE.9.002214

14. Merrill D, Sun H, Turek J, Nolte D, Yakubov B, Matei D, and An R, Intracellular Doppler Signatures of Platinum Sensitivity Captured by Biodynamic Profiling in Ovarian Xenografts, Nature Scientific Reports, 2016  doi.org/10.1038/srep18821

15. Nolte D, An R, Turek J, and Jeong K, Holographic tissue dynamics spectroscopy, *Journal of Biomedical Optics*, 2011 doi.org/10.1117/1.3615970

16. Choi H, Li Z, Hua Z, Zuponcic J, Ximenes E, Turek J, Ladisch M and Nolte D, Doppler imaging detects bacterial infection of living tissue, *Communications Biology, Nature*, 2021 doi.org/10.1038/s42003-020-01550-8